\pdfoutput=1
\documentclass[twocolumn,nofootinbib,preprintnumbers,amsmath,amssymb]{revtex4-1}
\usepackage{amssymb}
\usepackage{graphicx}
\usepackage{epsfig}

\begin{document}

\title{Implications of a 125 GeV Higgs for the MSSM and Low-Scale SUSY Breaking}
\author{Patrick Draper$^{1}$} 
\author{Patrick Meade$^{2}$}
\author{Matthew Reece$^{3}$}
\author{David Shih$^{4}$}

\affiliation{$^1$SCIPP, University of California, Santa Cruz, CA  95064\\
$^2$CNYITP, Stony Brook University, Stony Brook NY 11794\\
$^3$Department of Physics, Harvard University, Cambridge, MA 02138 \\
$^4$NHETC, Rutgers University, Piscataway, NJ 08854}

\date{\today}

\begin{abstract}
Recently, the ATLAS and CMS collaborations have announced exciting hints for a Standard Model-like Higgs boson at a mass of $\approx 125$ GeV. In this paper, we explore the potential consequences for the MSSM and low scale SUSY-breaking.  As is well-known, a 125 GeV Higgs implies either extremely heavy stops ($\gtrsim 10$ TeV), or near-maximal stop mixing. We review and quantify these statements, and investigate the implications for models of low-scale SUSY breaking such as gauge mediation where the $A$-terms are small at the messenger scale. For such models, we find that either a gaugino must be superheavy or the NLSP is long-lived. Furthermore, stops will be tachyonic at high scales. These are very strong restrictions on the mediation of supersymmetry breaking in the MSSM, and suggest that if the Higgs truly is at 125 GeV, viable models of gauge-mediated supersymmetry breaking are reduced to small corners of parameter space or must incorporate new Higgs-sector physics.
\end{abstract}

\maketitle

\section{Introduction}\label{motivation}

Recently, intriguing hints of the Standard Model (SM)-like Higgs boson have been reported by the LHC. The ATLAS collaboration has presented results in the diphoton~\cite{atlasgammagamma} and $ZZ^* \to 4\ell$~\cite{atlasZZ} channels, showing a combined $\sim 3\sigma$ excess at $m_h \approx 126$ GeV. The CMS collaboration has also presented results with a weaker $\sim 2\sigma$ excess in the $\gamma\gamma$ channel at $m_h \approx 123$ GeV~\cite{CMSgammagamma} and two events in the $ZZ^*$ channel near the same mass~\cite{CMSZZ}. It is too early to say whether these preliminary results will grow in significance to become a Higgs discovery, but it is not too early to consider some of the consequences if they do.

The potential discovery of a light Higgs renews the urgency of the gauge hierarchy problem. Supersymmetry remains the best-motivated solution to the hierarchy problem. Although it has not yet been found at the LHC, considerable discovery potential still remains in the parameter space relevant for naturalness~\cite{chickenfoot}. However, a 125 GeV Higgs places stringent constraints on supersymmetry, especially in the context of the minimal supersymmetric standard model (MSSM). In this paper we will examine these constraints in detail and use this to study the implications for low-scale SUSY breaking. 

In the MSSM, for values of the $CP$-odd Higgs mass $m_A\gtrsim 200$ GeV, there exists a light $CP$-even Higgs state in the spectrum with SM-like couplings to the electroweak gauge bosons. The SM-Higgs mass and properties are dominantly controlled by just a few weak-scale MSSM parameters: at tree level, $m_A$ and $\tan\beta$, joined at higher order by the stop masses $m_{\tilde t_{1,2}}$ and the stop mixing parameter $X_t\equiv A_t-\mu\cot\beta$. At tree-level, the Higgs mass is bounded above by $m_Z\cos2\beta$. One-loop corrections from stops are responsible for lifting this bound to $\sim130$ GeV~\cite{Haber:1990aw,Barbieri:1990ja,mhiggsRG1a,mhiggsRG1,HHH,Ambrosanio:2001xb}, for a general review, see~\cite{Carena:2002es}. Other parameters of the MSSM contribute radiative corrections to the Higgs mass, but in general are highly subdominant to the stop sector.  Even with large loop effects, it is noteworthy that 125 GeV is a relatively large Higgs mass for the MSSM---this fact allows us to constrain the stop masses and mixing.

In this paper, we will focus on stop masses $m_{\tilde t}\lesssim 5$ TeV which includes the collider relevant region. (We briefly consider heavier stops in the appendix.) Here  fixed-order Higgs spectrum calculators such as  FeynHiggs~\cite{Heinemeyer:1998yj,Heinemeyer:1998np,Degrassi:2002fi,Frank:2006yh}, which implements a broad set of one and two-loop corrections to the physical Higgs mass, should be fairly accurate.  Imposing an upper bound on the stop masses implies stringent bounds on $\tan\beta$ and $A_t$, and in particular requires large mixings among the stops.

For $m_A\lesssim 500$ GeV, the SM-like Higgs has an enhanced coupling to the down-type fermions, leading to an increase in the $h\rightarrow b\bar{b}$ partial width and suppressing the branching fractions into the main low-mass LHC search modes, $h\rightarrow\gamma\gamma,WW$~\cite{Carena:2011fc,Carena:1999bh,Cao:2011pg}. Since the LHC sees a rate consistent with SM expectations (albeit with a sizeable error bar), in this work we take $m_A=1$ TeV, where all the Higgs couplings are SM-like. This limit also avoids constraints from direct searches for $H/A\rightarrow\tau\tau$~\cite{Chatrchyan:2011nx,atlastautau,cmstautau}. For $\tan\beta$ we will set a benchmark value of 30 and consider a range of values in some cases.

\section{Implications for weak-scale MSSM parameters}

For $m_{\tilde t}\lesssim 5$ TeV, a Higgs mass of $m_{h}\approx 125$ GeV places strong constraints on $\tan\beta$ and the stop parameters. Although we will use FeynHiggs for all the plots in this section, it is useful to keep in mind the approximate one-loop formula for the Higgs mass,
\begin{eqnarray}
\label{mhtb}
&& m_h^2=m_Z^2c_{2\beta}^2\nonumber \\  && \quad + \frac{3m_t^4}{4\pi^2 v^2} \left(\log\left(\frac{M_S^2}{m_t^2}\right)+\frac{X_t^2}{M_S^2}\left(1-\frac{X_t^2}{12 M_S^2}\right)\right)
\end{eqnarray}
as it captures many of the qualitative features that we will see. We have characterized the scale of superpartner masses with $M_S \equiv \left(m_{\tilde t_1} m_{\tilde t_2}\right)^{1/2}$. First, we see that decreasing $\tan\beta$ always decreases the Higgs mass, independent of all the other parameters (keeping in mind that $\tan\beta\gtrsim 1.5$ for perturbativity). So we expect to find a lower bound on $\tan\beta$ coming from the Higgs mass. Second, we see that the Higgs mass depends on $X_t/M_S$ as a quartic polynomial, and in general it has two peaks at $X_t/M_S\approx \pm\sqrt{6}$, the ``maximal mixing scenario" \cite{HHH}. So we expect that $m_{h}=125$ GeV intersects this quartic in up to four places, leading to up to four preferred values for $X_t/M_S$. Finally, we see that for fixed $X_t/M_S$, the Higgs mass only increases logarithmically with $M_S$ itself. So we expect a mild lower bound on $M_S$ from $m_{h}=125$ GeV.

\begin{figure}[!t]
\includegraphics[width=0.45\textwidth]{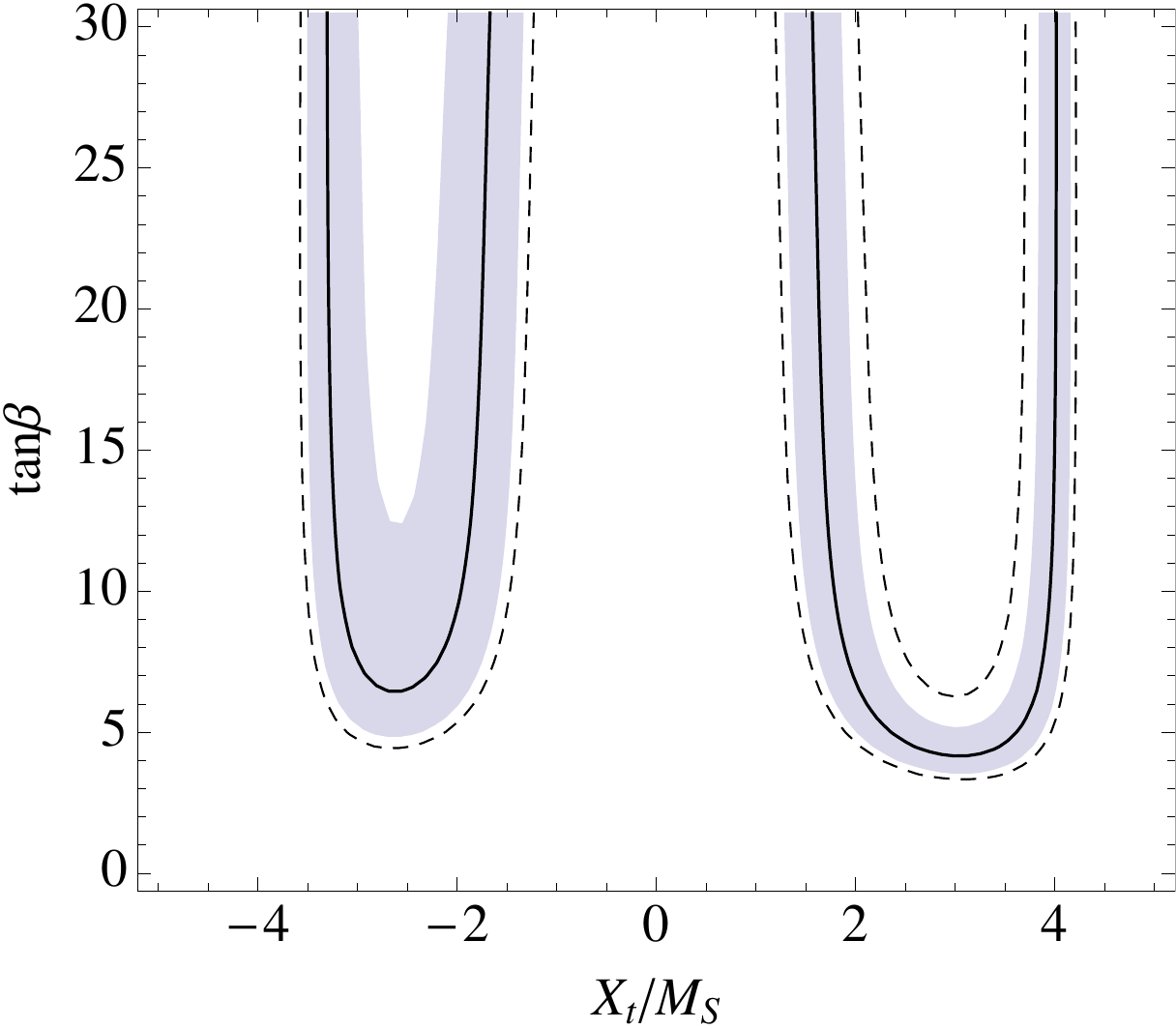}
\caption{Contour plot of $m_{h}$ in the $\tan\beta$ vs.\ $X_t/M_S$ plane. The stops were set at $m_Q=m_U=2$ TeV, and the result is only weakly dependent on the stop mass up to $\sim 5$ TeV. The solid curve is $m_{h}=125$ GeV with $m_t=173.2$ GeV. The band around the curve corresponds to $m_{h}=$123-127 GeV. Finally, the dashed lines correspond to varying $m_t$ from 172-174. }
\label{fig:tanbeta}
\end{figure}

Now let's demonstrate these general points with detailed calculations using FeynHiggs. Shown in fig.\ \ref{fig:tanbeta} are contours of constant Higgs mass in the $\tan\beta$, $X_t/M_{S}$ plane, for $m_Q=m_U=2$ TeV (where $m_Q$ and $m_U$ are the soft masses of the third-generation left-handed quark and right-handed up-type quark scalar fields). The shaded band corresponds to $m_{h}=123-127$ GeV, and the dashed lines indicate the same range of Higgs masses but with $m_t=172-174$ GeV. (The central value in all our plots will always be $m_{h}=125$ GeV at $m_t=173.2$ GeV.) From all this, we conclude that to be able to get $m_{h}\approx 125$ GeV, we must have
\begin{equation}
\tan\beta\gtrsim 3.5
\end{equation}
So this is an absolute lower bound on $\tan\beta$ just from the Higgs mass measurement. We also find that the Higgs mass basically ceases to depend on $\tan\beta$ for $\tan\beta$ beyond $\sim 20$. So for the rest of the paper we will take $\tan\beta=30$ for simplicity.

\begin{figure}[!t]
\begin{center}
\includegraphics[width=0.45\textwidth]{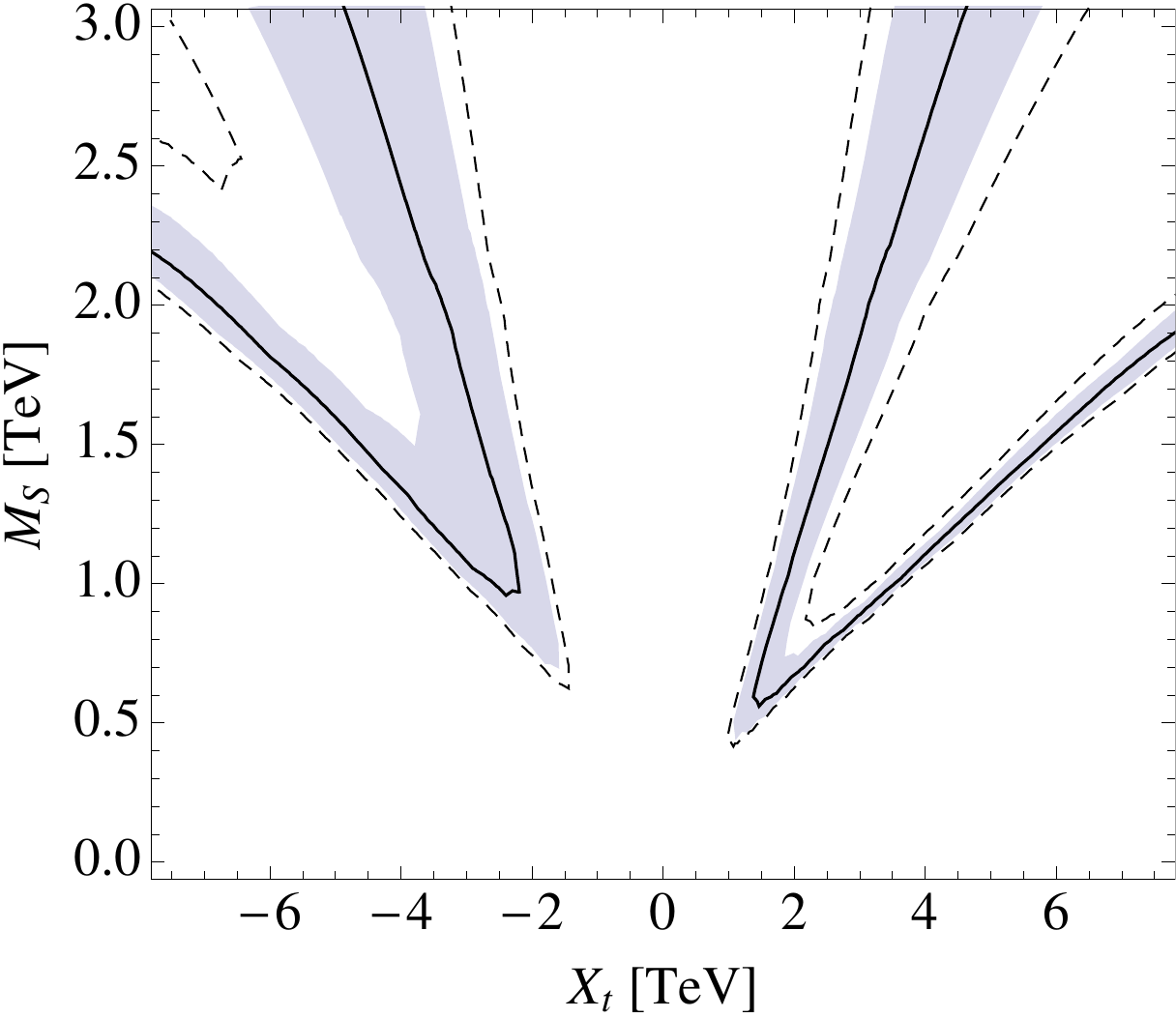}
\end{center}
\caption{Contours of constant $m_{h}$ in the $M_{S}$ vs.\ $X_t$ plane, with $\tan\beta=30$ and $m_Q=m_U$. The solid/dashed lines and gray bands  are as in fig.\ \ref{fig:tanbeta}. }
\label{fig:mhcurves}
\end{figure}

\begin{figure*}[t]
\includegraphics[width=0.9\textwidth]{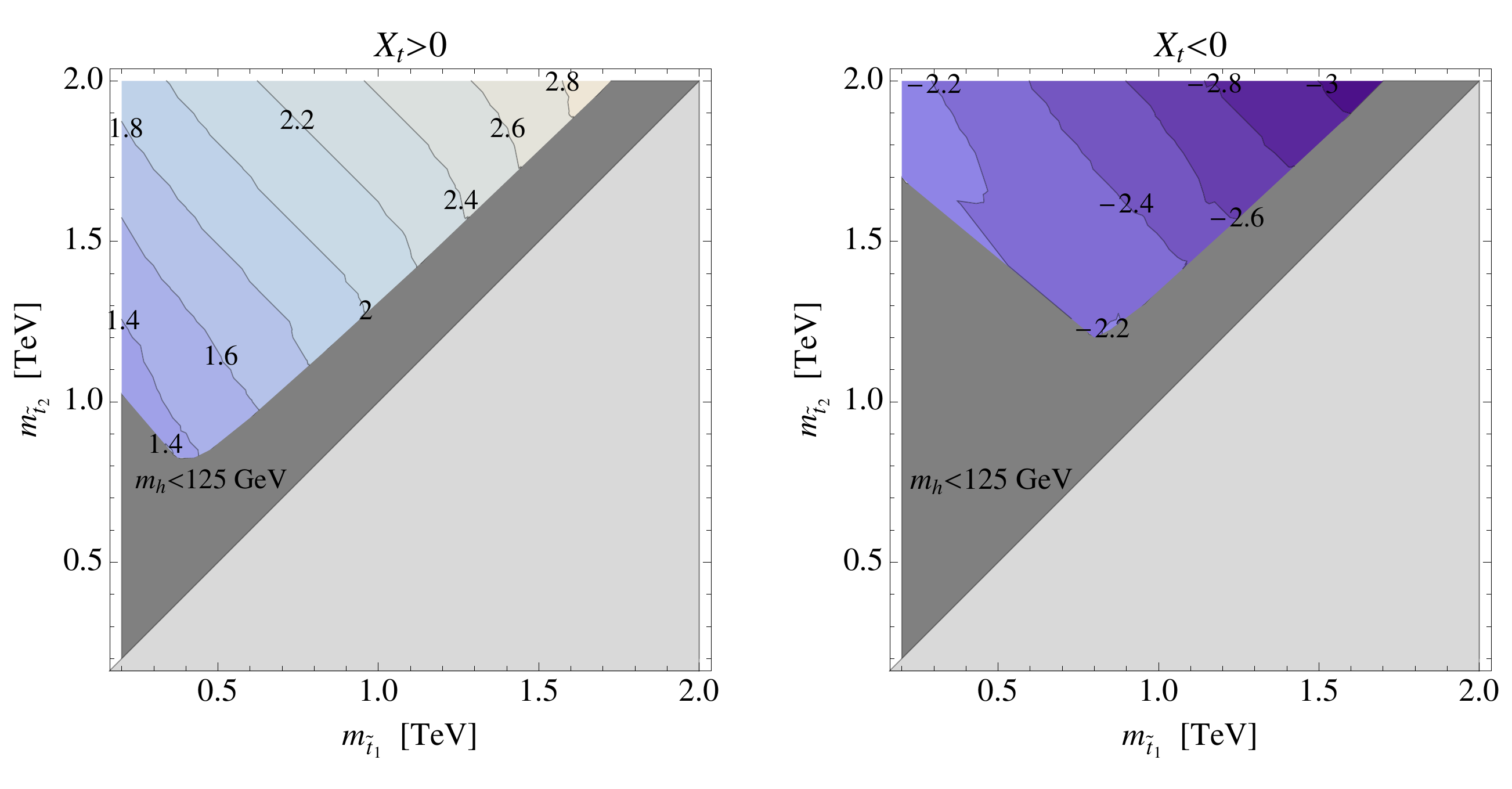}
\caption{Contour plot of $X_t$ in the plane of physical stop masses $(m_{\tilde t_1},\, m_{\tilde t_2})$. Here $X_t$ is fixed to be the absolute minimum positive (left) or negative (right) solution to $m_h=125$ GeV.}
\label{fig:mt1mt2At}
\end{figure*}

Fixing $\tan\beta$, the Higgs mass is then a function of $X_t$ and $M_S$. 
Shown in fig.\ \ref{fig:mhcurves} are contours of constant $m_h$ vs $M_S$ and $X_t$. We see that for large $M_S$, we want 
\begin{equation}
{X_t\over M_S}\approx -3,\,\,\,-1.7,\,\,\,1.5,\,\,\,{\rm or}\,\,\,3.5
\end{equation}
We also see that the smallest the $A$-terms and the SUSY-scale can absolutely be are 
\begin{equation}
|X_t| \gtrsim 1000\,\,{\rm GeV} ,\qquad M_S \gtrsim 500 \,\,{\rm GeV}.
\end{equation}
It is also interesting to examine the limits in the plane of physical stop masses. Shown in fig.\ \ref{fig:mt1mt2At} are plots of the contours of constant $X_t$ in the $m_{\tilde t_2}$ vs.\ $m_{\tilde t_1}$ plane. Here the values of $X_t<0$ and $X_t>0$ were chosen to satisfy $m_{h}=125$ GeV, and the solution with smaller absolute value was chosen. In the dark gray shaded region, no solution to $m_{h}=125$ GeV was found. Here we see that the $\tilde t_1$ can be as light as 200 GeV, provided we take $\tilde t_2$ to be heavy enough. We also see that the heavy stop has to be much heavier in general in the $X_t<0$ case.

\section{Implications for the SUSY Breaking Scale}

\begin{figure*}[t]
\includegraphics[width=0.85\textwidth]{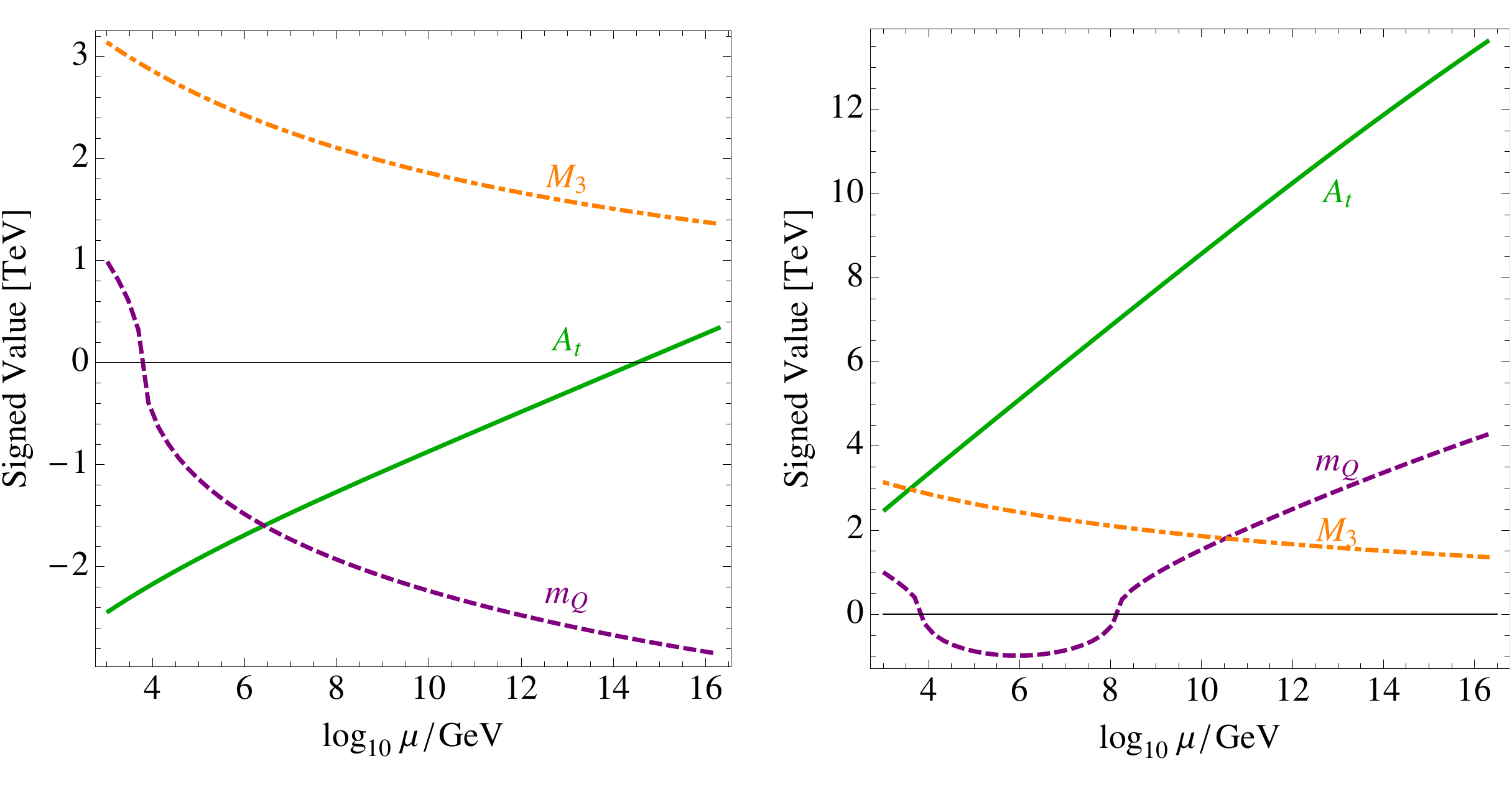}
\caption{Values of running parameters: at left, in a case where $A_t$ is large and negative at low scales; at right, in a case where it is large and positive. The case $A_t < 0$ at low scales can be compatible with $A_t = 0$ from a high-scale mediation scheme, and in this case we expect that it is generally associated with tachyonic squarks at a high scale. Scalar masses are plotted as signed parameters, e.g. $m_Q^{(plotted)} \equiv m_Q^2/\left|m_Q\right|$.}
\label{fig:RGEruns}
\end{figure*}

Having understood what $m_{h}\approx125$ GeV implies for the weak-scale MSSM parameters, we now turn to the implications for the underlying model of  SUSY-breaking and mediation. In RG running down from a high scale, for positive gluino mass $M_3$, the $A$-term $A_t$ decreases. The gluino mass also drives squark mass-squareds larger at small scales, whereas the $A$-term drives them smaller. The interplay among these effects is illustrated in the running of two sample spectra in Figure~\ref{fig:RGEruns}. We see that for negative $A_t$ at the weak-scale, RG running can drive $A_t$ across $A_t=0$ at some high scale, but for positive $A_t$ at the weak scale, RG running generally drives $A_t$ even higher.

This has important consequences for models of gauge mediated SUSY breaking (GMSB). (For a review and original references, see \cite{Giudice:1998bp}.) In pure gauge mediation (as defined e.g.\ in \cite{Meade:2008wd}), the $A$-terms are strictly zero at the messenger scale. This conclusion remains robust even when a sector is added to generate $\mu$/$B\mu$ \cite{Komargodski:2008ax}. Clearly, in models of GMSB with vanishing $A$-terms at the messenger scale, to get sufficiently large $A$ terms  at the weak scale, we must either run for a long time, or have heavy gauginos. 

\begin{figure*}[!t]
\begin{minipage}{.9\linewidth}
\begin{center}
\begin{tabular}{cc}
\includegraphics[width=0.45\textwidth]{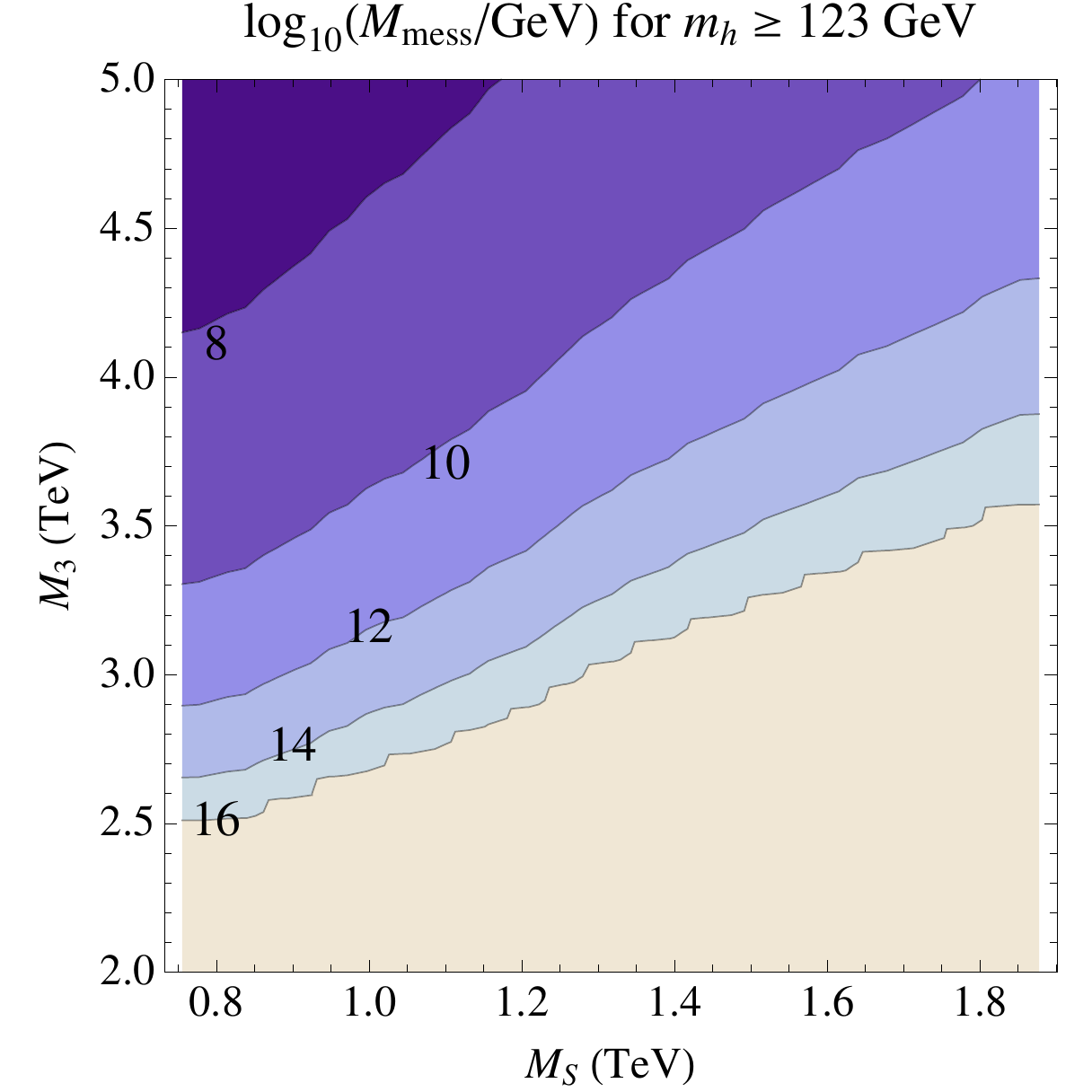} &
\includegraphics[width=0.45\textwidth]{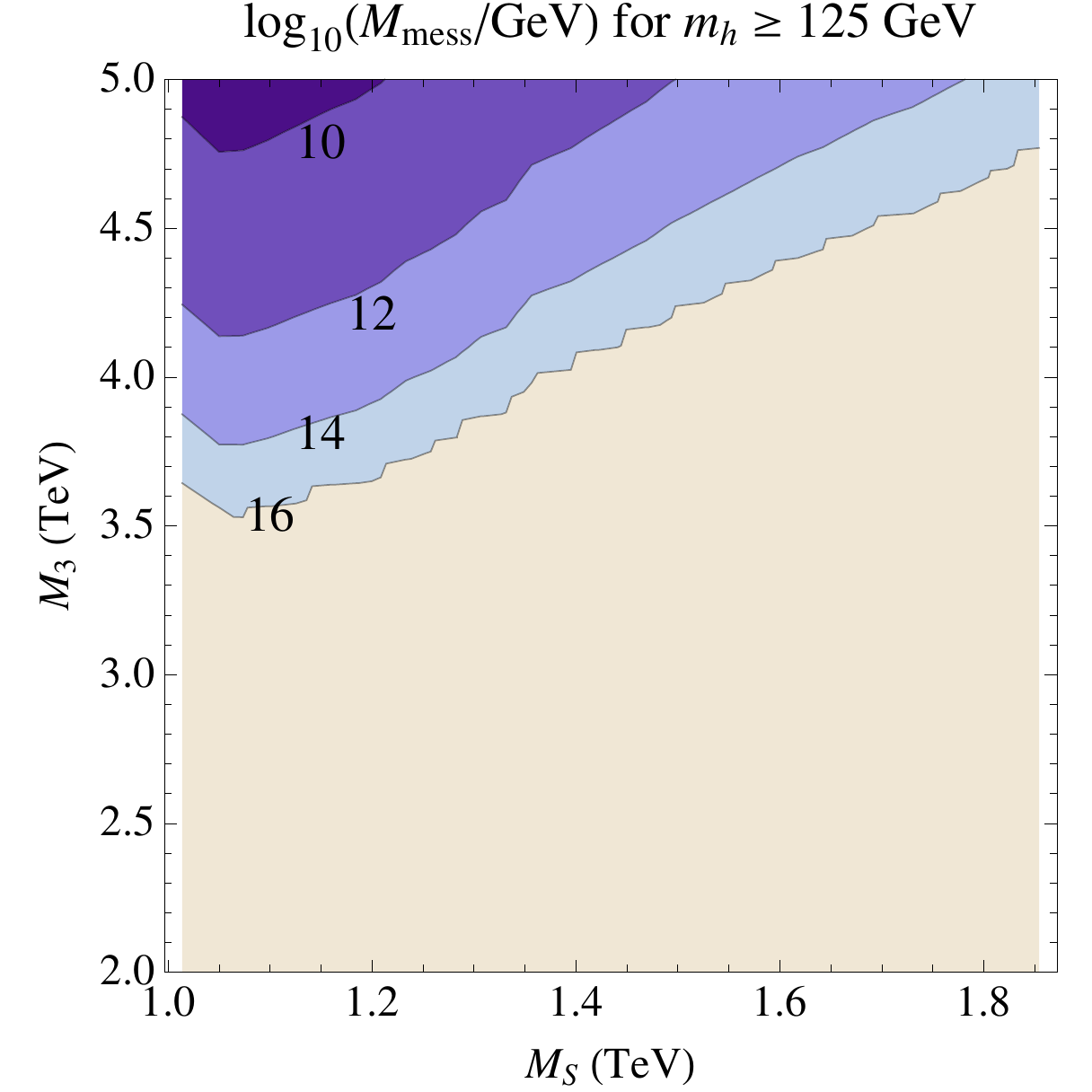}
\end{tabular}
\caption{Messenger scale required to produce sufficiently large $|A_t|$ for $m_h=123$ GeV (left) and $m_h=125$ GeV (right) through renormalization group evolution.}
\label{fig:mmess}
\end{center}
\end{minipage}
\end{figure*}

We can quantify this by starting from the EW scale ($m_{\tilde t_1}$, $m_{\tilde t_2}$) points that can produce the required Higgs mass with a negative value for $A_t$, and define a gauge mediation messenger scale by evolving $A_t$ until it vanishes.  In Fig.~\ref{fig:mmess} we plot the messenger scales obtained over the $(M_3, M_S)$ plane from this algorithm, performing the evolution with 1-loop $\beta$ functions, decoupling the gluino below $M_3$, and decoupling the rest of the superpartner spectrum at $M_S$.  Due to the logarithmic sensitivity of $A_t$ on the messenger scale, and the polynomial dependence of $m_h$ on $A_t$, the reconstructed messenger scale varies strongly between $m_h=123$ GeV and $m_h=125$ GeV, as shown in the two panels of Fig.~\ref{fig:mmess}. We see that large gluino masses are required to keep the messenger scale below the GUT scale. Notice that we have assumed the running of $A_t$ is dominated by $M_3$. If the superpartner spectrum had $M_2$ or $M_1 \gg M_3$, this conclusion might be avoided, but this requires unusual hierarchies in the gaugino spectrum.

In gauge mediation, the messenger mass scale is closely related to the lifetime for the decay of the next-to-lightest superpartner (NLSP) to the gravitino. We can quantify this as follows. Achieving superpartners in the TeV mass regime requires $F/M_{\rm mess} \sim 100$ TeV. A typical lifetime is determined by the coupling to the Goldstino, which is $1/F$ suppressed, and hence
\begin{equation}
c\tau_{NLSP} \sim \frac{16 \pi^2 F^2}{m_{NLSP}^5} \approx 3~{\rm m}~\left(\frac{M_{\rm mess}}{10^7~{\rm GeV}}\right)^2 \left(\frac{100~{\rm GeV}}{m_{NLSP}}\right)^5.
\end{equation}
Thus, a 10$^7$ GeV messenger scale corresponds to ``collider-size" decay lengths, and higher messenger scales correspond to ``collider-stable" NLSPs. Based on Figure~\ref{fig:mmess}, we see that achieving the requisite $A$-terms in gauge mediation with superpartners of order a TeV is only possible in the limit of long-lived or collider-stable NLSPs. Prompt decays (corresponding to much lower messenger scales) are allowed in GMSB (with vanishing $A$-terms at the messenger scale) only for extremely heavy gluinos, well above 4 TeV. For GMSB models with gluinos of a few TeV, messenger scales on the order of 10$^8$ GeV can exist, allowing for collider-size but not truly prompt decays. Furthermore, for $M_{\rm mess} \gtrsim 10^{11}$ GeV, the NLSP lifetimes become a second or longer, and NLSP decays can play a significant role in post-BBN cosmology~\cite{Moroi:1993mb,Gherghetta:1998tq}.  Additionally, gravitinos can overclose the universe if the reheating scale is too large, although this constraint is milder at large $M_{\rm mess}$~\cite{Moroi:1993mb}.

The squark soft masses squared are also strongly influenced by $M_3$ dependence in their $\beta$-functions. Typically they are driven negative at the messenger scale.
While this clearly requires non-minimal gauge mediation, it is not intrinsically problematic for the theory, and can in fact lead to improvements in the fine-tuning problems of the MSSM~\cite{Dermisek:2006ey}. Tachyonic stops at a high scale will lead to charge and color breaking (CCB) vacua, but the lifetime of our vacuum is typically long enough that the theory is not excluded~\cite{Riotto:1995am, Kusenko:1996jn}.  It is also important to make sure that our cosmological history is not affected by the presence of CCB vacua~\cite{Carena:2008vj}. Given that the CCB vacuum, long-lived NLSPs, and thermal production of gravitinos can all pose different constraints on the theory, it is possible that reheating in such a scenario is strongly constrained, and this deserves further study. It is also interesting, given relatively large messenger scales and hence gravitino masses, to consider the possibility of superWIMP gravitino dark matter~\cite{Feng:2003xh}.

\section{Conclusions}

Broadly speaking, the possible supersymmetric interpretations of a 125 GeV Higgs fall into three classes: large $A$-terms, heavy scalars, or modifications of the MSSM (such as extensions of the Higgs sector, compositeness, or new gauge groups). Within the MSSM, we have seen that restricting to light scalars requires an $A$-term above 1 TeV; avoiding the conclusion of large $A$-terms is only possible by decoupling scalars to at least 5 TeV. Thus, a Higgs at this mass is in some tension with MSSM naturalness.

Furthermore, the requirement of large $A$-terms is a powerful constraint on models of the supersymmetry breaking mechanism. Most notably, we have seen that there is a severe restriction on the allowed messenger scale in gauge mediation if superpartners are within reach at the LHC. This can be avoided only by extending the definition of gauge mediation, e.g. to include new Higgs-sector couplings to the SUSY breaking sector.

Looking forward, let us briefly comment on the Higgs cross section and branching ratio in the SM vs.\ the MSSM. The production of the Higgs at the LHC through gluon fusion and the Higgs decay to two photons are both loop-level processes in the Standard Model and can be sensitive to the presence of new physics~\cite{Djouadi:1998az,Djouadi:1996pb,Dermisek:2007fi}. However, as we have shown here, $m_h=125$ GeV generally requires quite heavy stops, and this limits the effect they can have on the cross section and branching ratio. Generally, we find a suppression of $\sigma\times {\rm Br}(\gamma\gamma)$ of at most 15\% through varying $m_{\tilde t_{1,2}}$ and requiring $m_h\approx 125$ GeV. We find much smaller suppressions of the effective cross sections into $WW$ and $ZZ$ given that they are tree level decays and less sensitive to higher scale physics. Whether these small deviations from the SM could be measured at the LHC is a subject for future study. Certainly we do not expect experimental results along these lines any time soon.

We should also comment on the assumptions that have gone into the conclusions here. First, we assumed the pure MSSM. This led directly to the large $A_t$ and heavy stop conclusions. If we relax this assumption (e.g.\ to the NMSSM), then many of these conclusions could change qualitatively. Second, for gauge mediation we assumed $A_t=0$ at the messenger scale. Clearly this is not completely set in stone, and it would be interesting to look for models of GMSB (or more generally flavor-blind models) with large $A_t$ at the messenger scale. This may be possible in more extended models, for instance in~\cite{Evans:2011bea} where the Higgses mix with doublet messengers. 

\acknowledgments{We would like to thank Michael Dine, Jared Evans, Howie Haber, Yevgeny Kats and Chang-Soon Park for useful discussions. PD is supported by the DOE under grant DE-FG02-04ER41286. The work of PM is supported in part by NSF CAREER Award NSF-PHY-1056833. MR is supported by the Fundamental Laws Initiative of the Center for the Fundamental Laws of Nature, Harvard University. The research of DS was supported in part by a DOE Early Career Award.
}

\appendix

\section{Comments on ``heavy SUSY" scenarios}
\label{sec:heavySUSY}

\begin{figure}[b]
\includegraphics[width=0.38\textwidth]{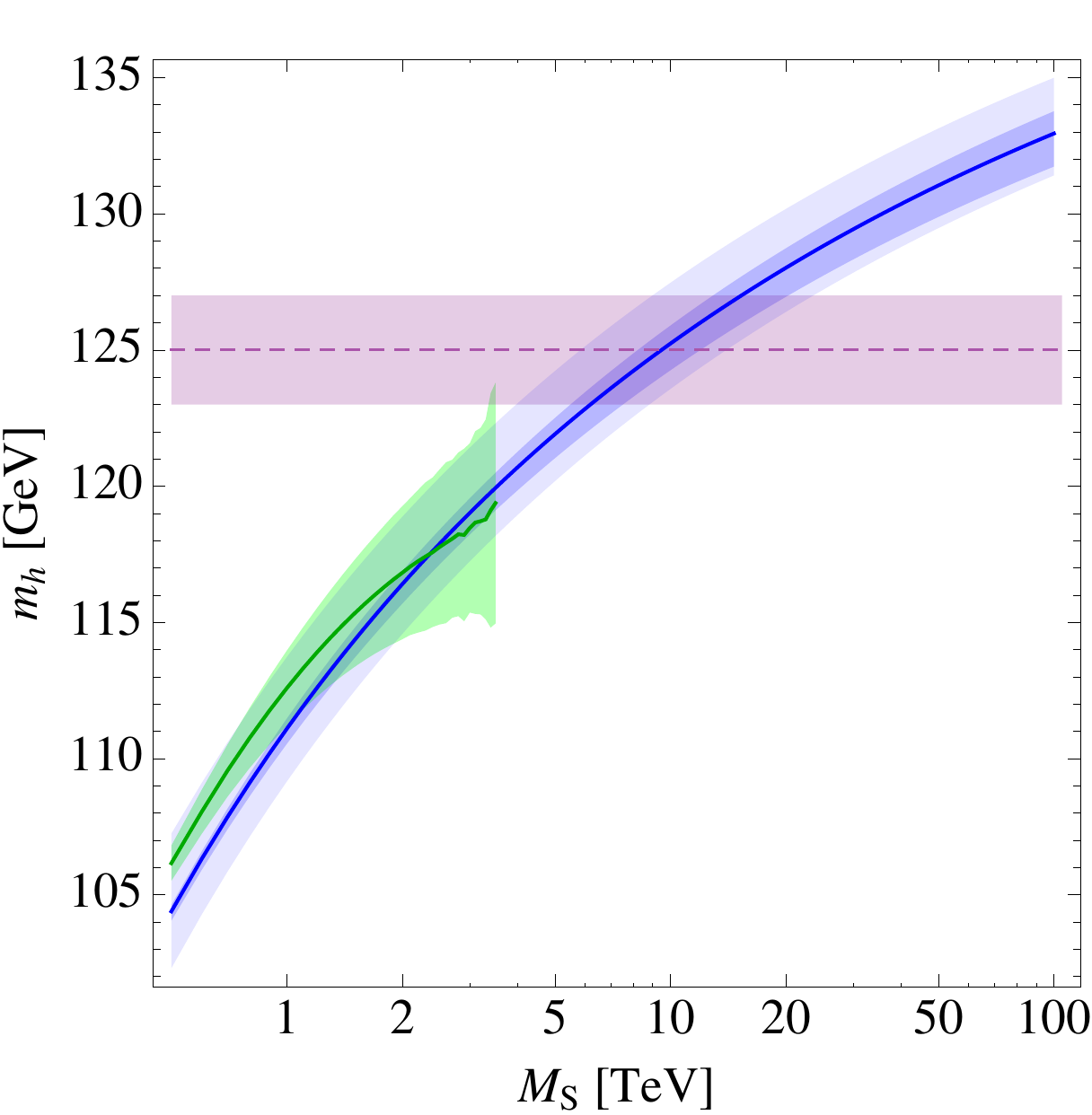}
\caption{Higgs mass as a function of $M_S$, with $X_t = 0$. The green band is the output of FeynHiggs together with its associated uncertainty. The blue line represents 1-loop renormalization group evolution in the Standard Model matched to the MSSM at $M_S$. The blue bands give estimates of errors from varying the top mass between 172 and 174 GeV (darker band) and the renormalization scale between $m_t/2$ and $2 m_t$ (lighter band).}
\label{fig:highscalemh}
\end{figure}

Although we have focused on mixed stops which can be light enough to be produced at the LHC, let us briefly consider the case of stops {\em without} mixing. For small $M_S$, we can compute the Higgs mass with FeynHiggs. For larger $M_S$, we use a one-loop RGE to evolve the SUSY quartic down to the electroweak scale, computing the physical Higgs mass by including self-energy corrections~\cite{Sirlin:1985ux,Hempfling:1994ar}. In Figure~\ref{fig:highscalemh}, we plot the resulting value of $m_h$ as a function of $M_S$, in the case of zero mixing. We plot the FeynHiggs output only up to 3 TeV, at which point its uncertainties become large and the RGE is more trustworthy. One can see from the plot that accommodating a 125 GeV Higgs in the MSSM with small $A$-terms requires scalar masses in the range of 5 to 10 TeV. 

A variation on this ``heavy stop" scenario is Split Supersymmetry~\cite{ArkaniHamed:2004fb,Arvanitaki:2004eu}, in which gauginos and higgsinos have masses well below $M_S$ and influence the running of $\lambda$. In this case, the running below $M_S$ is modified by the light superpartners, and the preferred scalar mass scale for a 125 GeV Higgs can be even larger~\cite{Giudice:2004tc,Bernal:2007uv,Giudice:2011cg}.

\end{document}